\documentclass{aa}
\usepackage{graphicx}
\usepackage{txfonts}

\usepackage{natbib} 
\bibpunct{(}{)}{;}{a}{}{,}

\begin{document}

 \title{A 3-D sunspot model derived from an inversion of spectropolarimetric observations and its implications for the penumbral heating}

   \author{C. Beck\inst{1,2} }
        
   \titlerunning{A 3-D sunspot model}
  \authorrunning{C. Beck}  
\offprints{C. Beck}

   \institute{Instituto de Astrof\'{\i}sica de Canarias
     (CSIC), La Laguna, Tenerife, Spain\\
      \and Kiepenheuer-Institut f\"ur Sonnenphysik,
  , Freiburg, Germany.\\
    }
 
\date{Received xxx; accepted xxx}

\abstract{}{I deduced a three-dimensional sunspot model that is in full agreement with spectropolarimetric observations, in order to address the question of a possible penumbral heating process by the repetitive rise of hot flow channels.}{I performed inversions of spectropolarimetric data taken
  simultaneously in infrared (1.5\thinspace $\mu$m) and visible (630\thinspace
  nm) spectral lines. I used two independent magnetic components inside each
  pixel to reproduce the irregular Stokes profiles in the penumbra. I studied
  the averaged and individual properties of the two components. By integrating
  the field inclination to the surface, I developed a three-dimensional model
  of the spot from inversion results without intrinsic height
  information.}{I find that the Evershed flow is harbored by the weaker of
  the two field components. This component forms flow channels that show 
  upstreams in the inner and mid penumbra, continue almost horizontally as slightly elevated loops throughout the penumbra, and finally bend down in the outer
  penumbra. I find several examples, where two or more flow channels are found along a radial cut from the umbra to
  the outer boundary of the spot.}{I find that a model of horizontal flow channels in a static
  background field is in good agreement with the observed spectra. The
  properties of the flow channels correspond very well to the Moving Tube
  simulations of Schlichenmaier et al. (1998). From the temporal evolution in intensity images and the properties of the flow channels in the inversion, I
  conclude that interchange convection of rising hot flux tubes in a thick
  penumbra still seems a possible mechanism for maintaining the penumbral energy balance.}
\keywords{Sun: magnetic fields, Sun: sunspots} 
\maketitle
\begin{figure}
\centerline{\resizebox{7cm}{!}{\includegraphics{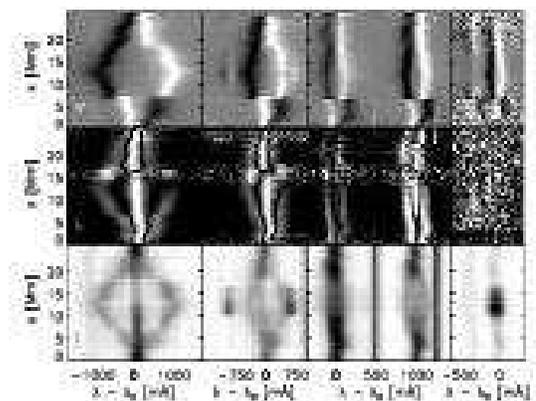}}}
\caption{Stokes profiles along the symmetry line connecting spot center and
  disc center. The limb side is down, the center side up. {\em Bottom row}:
  intensity, $I$. {\em Middle row}: Total linear polarization, $L$. {\em Top
    row}: Circular polarization, $V$. {\em Left to right}: IR spectral lines at 1564.8\thinspace nm, 1565.2\thinspace nm, VIS lines at 630.15\thinspace nm, 630.25\thinspace nm, and \ion{Ti}{I} at 630.37\thinspace nm.\label{linpol} }
\end{figure}
\section{\label{sec_int}Introduction}
Sunspots have been the first indicators of solar magnetic activity detected
already centuries ago in the Western World
\citep{galilei+etal1613,galilei1632}, respectively, millennia ago in Asia
\citep{wittmann+xu1987}. Their magnetic nature was only proven in
the last century by \citet{hale1908}. Using spectroscopic observations,
\citet{evershed1909} showed that sunspots exhibit a particular
flow field in the penumbra, the brighter halo surrounding the dark umbra: as
soon as a sunspot was observed not directly on disc center, spectral lines in
the half of the spot facing disc center were seen to be blue shifted,
whereas they were red shifted in the half facing the limb. This ``Evershed
flow'' was explained as the signature of a radial outflow, which had to be
close to parallel to the solar surface. 

Despite its rather well defined spectroscopic properties, the Evershed flow however still eludes a generally
accepted explanation. The reason is the fine-structure of the sunspot's
penumbra that became visible when both the spatial resolution and the
information content of observations was improved. Spectroscopic observations
with high spatial resolution show that the Evershed flow is organized in
many small-scale radially oriented filaments, which carry the bulk of the
flow, whereas in the space in-between the flow velocity is strongly reduced
\citep{tritschler+etal2004,langhans+etal2005,rimmele+marino2006}. Different,
and in some cases, contradictory results were found on the correlation between
these flow filaments and intensity filaments \citep[cf.~the discussion
in][]{langhans+etal2005}. 
\begin{figure*}
\centerline{\resizebox{17.6cm}{!}{\includegraphics{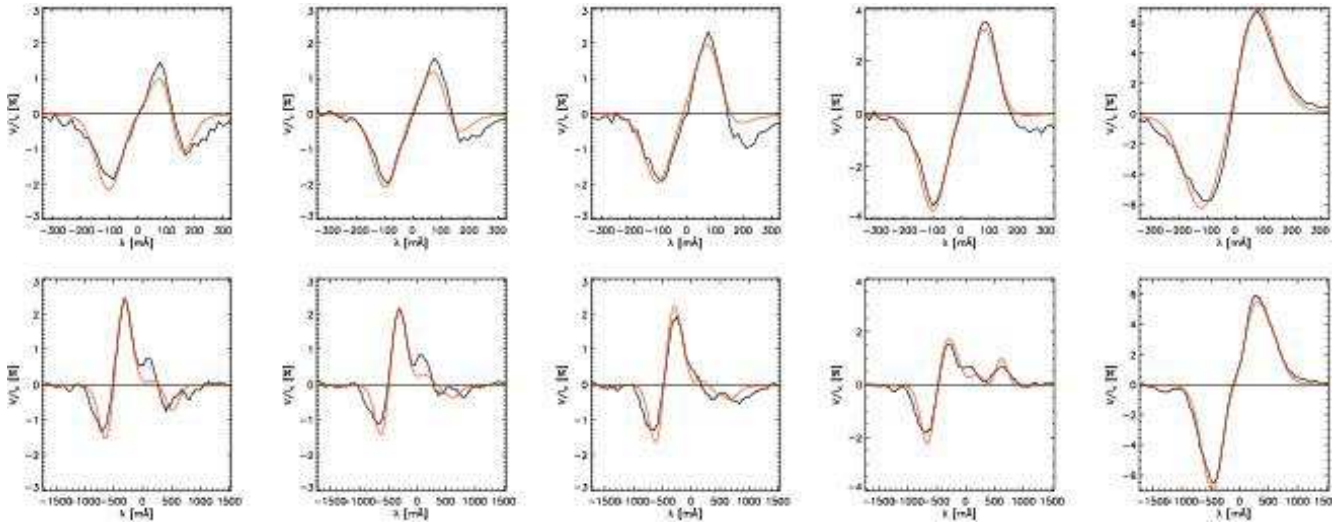}}}
\caption{Examples of Stokes $V$ profiles in the neutral line. {\em Top row}:
  Stokes $V$ of the \ion{Fe}{I} line at 630.15\thinspace nm.  {\em Bottom
    row}: The co-spatial profiles of \ion{Fe}{I} at 1564.8\thinspace nm. The
  red line gives the corresponding best-fit profile of the
  inversion. The last column shows a profile in the center side
  penumbra.\label{vneutral}}
\end{figure*}

Using spectropolarimetric observations of sunspots, the information content
of the observations was extended to encompass both the magnetic field topology
and the flow fields at the same time. The problem of the penumbral fine
structure however was not resolved by this, but got more complex again. The
Evershed flow could be shown to be close to horizontal, but the {\em average}
magnetic field turned out not to be
\citep[e.g.][]{bellot+balthasar+etal2003}. A possible explanation for the mismatch between flow and field direction in an ionized
plasma is the assumption that at the spatial resolution of around
1$^{\prime\prime}$, achieved in current spectropolarimetric observations, two
different magnetic field components are seen inside the same pixel. These two
magnetic field components have a different topology and flow field each. 

One way to visualize this is to take a cut along the
symmetry line of a sunspot, the line that passes from disc center across the
spot center towards the closest solar limb position \citep{collados2002,bellot+etal2004}. Figure \ref{linpol} shows
profiles of intensity, total linear polarization, $L = \sqrt{Q^2+U^2}$, and
circular polarization (Stokes $V$) along the symmetry line for one of the observations analyzed in the
present study. Both circular and linear polarization show a neutral line on
the limb side and center side, respectively. At the location of these neutral
lines, the profile shape changes abruptly. On the limb side across the neutral
line of Stokes $V$, stronger fields with smaller line shifts turn into weaker
fields with larger line shifts. On the center side across the neutral line of
$L$, the same change from strong fields with slower flows to weaker fields
with stronger flows is seen.

A more direct evidence of these multiple magnetic field components is given by
the Stokes $V$ profiles close to or in the neutral line of $V$
(Fig.~\ref{vneutral}) on the limb side. The neutral line indicates a local minimum of the
  Stokes $V$ signal amplitude, because most of the field lines are
  perpendicular to the line of sight. In magnetograms, the retrieved signal
  passes through zero in the neutral line, because the polarity of the
  magnetic field changes from, e.g.~positive to negative, but in spectra the
  $V$ signal does not disappear completely due to the multi-component
  penumbral structure \citep[cf.][]{schlichenmaier+collados2002}. These profiles  in the neutral line show not only two $\sigma$-components in
the circular polarization signal, but several local extrema (both minima and
maxima). To generate such complex patterns, the magnetic field has to be
complex as well. Appendix \ref{prof_constr} shows how such complex profiles can be created by the addition of two regular Stokes $V$ profiles with opposite polarities. This simple addition of profiles however does not tell, if the
two magnetic field components leading to the $V$ profiles are actually located in the same pixel. If the spatial resolution of the polarimetric observations is insufficient to separate the flow filaments seen in high resolution observations from their surroundings, field components from two different spatial locations would be added up. 

There however is an indicator that suggests that the field components are not spatially separated in the
horizontal dimensions, but vertically interlaced, the Net Circular
Polarization (NCP). A non-zero value of NCP -- as observed in the penumbra of
sunspots
\citep[e.g.][]{almeida+lites1992,solanki+montavon1993,valentin2000,mueller+etal2002,mueller+etal2006} --
indicates discontinuities or at least strong gradients of
field properties along the line of sight inside the formation height of
spectral lines. 

Most of the observed properties of profiles in the penumbra of sunspots can be
 explained with the {\em uncombed} penumbra model suggested by
 \citet{solanki+montavon1993}. In this model, a more vertical field component
 winds around horizontal flow channels, which carry the Evershed flow. The success of this model is related to the fact that
 it is able to explain the most prominent peculiarities of sunspots: a
 horizontal flow field in a non-horizontal field topology, the
 appearances of neutral lines in total linear polarization, $L$, and Stokes
 $V$, and the NCP. However, the model itself gives no reason, why its specific topology should exist in the
 penumbra. The simulations of \citet{schliche+jahn+schmidt1998} showed that
 such a configuration of embedded flow channels results from the temporal
 evolution of flux initially located at the boundary layer between the sunspot
 and the field-free surroundings. \citet{thomas+weiss2004} suggested that
 turbulent convection outside the spot pulls down field lines and thus
 produces the filamentary structure of the sunspot. In both cases, the
 penumbra consists of similar embedded flow channels, but created by different
 mechanisms. \citet{spruit+scharmer2006} and \citet{scharmer+spruit2006}
 recently suggested a completely different model, where the intensity
 filaments are related to the existence of field-free gaps below the visible
 surface layer. 

 In this paper, I want to show that a rather simple uncombed model
   using two depth-independent magnetic components is  {\em sufficient}
   to reproduce simultaneous observations in infrared and visible spectral
 lines with their different responses to magnetic fields. The inversion setup is described after the observations and data reduction (Sect.~\ref{sec_obs}). I study average field properties and their radial variation in Sect.~\ref{sec_results}. I then deduce a geometrical model of the sunspot by integrating the surface inclination of the fields (Sect.~\ref{sec_3dmodel}). The physical properties of the model like field strength, velocity, or field orientation, correspond to the best-fit model atmospheres necessary to reproduce the observed spectra. I shortly investigate the temporal evolution in Sect.~\ref{tempevol}, with emphasis on which parts change with time and which do not. In the discussion (Sect.~\ref{sec_disc}), I address the question of penumbral heat transport in the context of the results of the previous analysis. Appendix \ref{prof_constr} shows how complex profiles can be created with simple assumptions. Appendix \ref{appa} shows an example of the integration along a single cut through the penumbra and the full 3-D model from different viewing angles. Appendix \ref{appb} describes the analysis of a time series of speckle-reconstructed G-band images used to derive the characteristic time scale of penumbral intensity variations.

I used a simple inversion setup with field properties constant
with optical depth, corresponding to a horizontal separation of the field
components. A more sophisticated setup, taking into account the variations along the line-of-sight, is planned to be discussed in a forthcoming paper.
\begin{figure}
\centerline{\resizebox{8.cm}{!}{\includegraphics{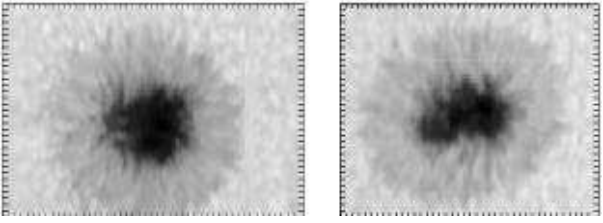}}}
\caption{Continuum intensity maps of NOAA 10425 in the near infrared on
  7.8.2003 ({\it left}) and 9.8.2003 ({\it right}). The distance between tick marks is 1 arcsec. \label{spotim}}
\end{figure}
\section{\label{sec_obs}Observations \& Data reduction}
For this work, two observations of the sunspot NOAA 10425 were analyzed. They were taken on the 7th and 9th of August 2003 with the two vector
spectropolarimeters of the German Vacuum Tower Telescope (VTT) on
Tenerife: the Polarimetric Littrow Spectrograph
\citep[POLIS;][]{beck+etal2005b} and the Tenerife Infrared Polarimeter \citep[TIP;][]{martinez+etal1999}. The two instruments were used
simultaneously using an achromatic 50-50-beamsplitter. During the observations, POLIS was remote-controlled by TIP to ensure
strictly simultaneous exposures. The image motion was reduced using the
Correlation Tracker System \citep{schmidt+kentischer1995,ballesteros+etal1996}. The Stokes vector in the visible and infrared spectral lines listed in Table \ref{tablines} was retrieved.
\begin{table}
\caption{Spectral lines observed with  POLIS ({\em top}) and TIP ({\em
    bottom}).\label{tablines}}
\begin{center}
\begin{tabular}{lcccc} \hline\hline  
element & $\lambda$ & transition & log ($gf$)& Land{\'e} factor  \cr  
ion. state & $[$nm$]$ & $^{2S+1}L_{J}$ & & $g_{eff}$  \cr \hline 
\multicolumn{5}{c}{\rule[-2mm]{0mm}{6mm} {\bf POLIS}} \cr \hline 
\ion{Fe}{I} & 630.15012$^a$ & $^5 P_{2} - ^5\hspace*{-.1cm} D_{2} $& -0.75$^d$ & 1.67 \cr 
\ion{Fe}{I} & 630.24936$^a$ & $^5 P_{1} - ^5 \hspace*{-.1cm}D_{0} $& -1.236$^b$& 2.50\cr
\ion{Fe}{I}$^b$ & 630.34600 & $^5 G_{6} - ^5 \hspace*{-.1cm}G_{5} $& -2.55&  1.50\cr
\ion{Ti}{I}$^b$ & 630.37525 &  $^3 F_{3} - ^3\hspace*{-.1cm} G_{3}$ & -1.44&0.92\cr 
\multicolumn{5}{c}{\rule[-2mm]{0mm}{6mm} {\bf TIP}} \cr \hline 
\ion{Fe}{I}$^c$  & 1564.7410 & $^7 D_{2} - ^5\hspace*{-.1cm} P_{2}$ & -0.95&1.25 \cr
\ion{Fe}{I}$^c$  & 1564.8515 & $^7 D_{1} - ^7\hspace*{-.1cm} D_{1}$ & -0.67& 3 \cr
\ion{Fe}{I}$^c$  & 1565.2874 & $^7 D_{5} - ^7\hspace*{-.1cm} D_{4}$ & -0.095$^d$&1.45 \cr\hline
\end{tabular}\\
(a) \citet{nave+etal1994}; (b) \citet{cabrera+bellot+iniesta2005}\\ (c)
\citet{bellot+etal2000}; (d) L.~R.~Bellot Rubio, priv. comm.\\ All values are
given as used in the inversion; they may slightly deviate from the sources
cited in $\lambda$ and log($gf$) due to changes in the adopted solar iron abundance and new IR line measuerements.
\end{center}
\end{table}
An integration time of around 3.5 sec was used for each slit position. The slit
width was 0\farcs36 for TIP and 0\farcs5 for POLIS. The scanning
step width was 0\farcs36. Spatial sampling along the slit was 0\farcs37 for TIP
and 0\farcs145 for POLIS. The POLIS data were interpolated later to have the
same spatial sampling along the slit as TIP. Figure \ref{spotim} displays maps of the continuum intensity in the infrared. The 2-D intensity maps were constructed from the
intensity along the slit for each scan step, the scanning direction is left to
right. The  heliocentric angle of the spot on the August 7th and 9th was
7$^\circ$ and 30$^\circ$, respectively. The spatial resolution was estimated from the spatial Fourier power spectrum to be around 1$^{\prime\prime}$.
\begin{figure*}
\centerline{\resizebox{17.6cm}{!}{\includegraphics{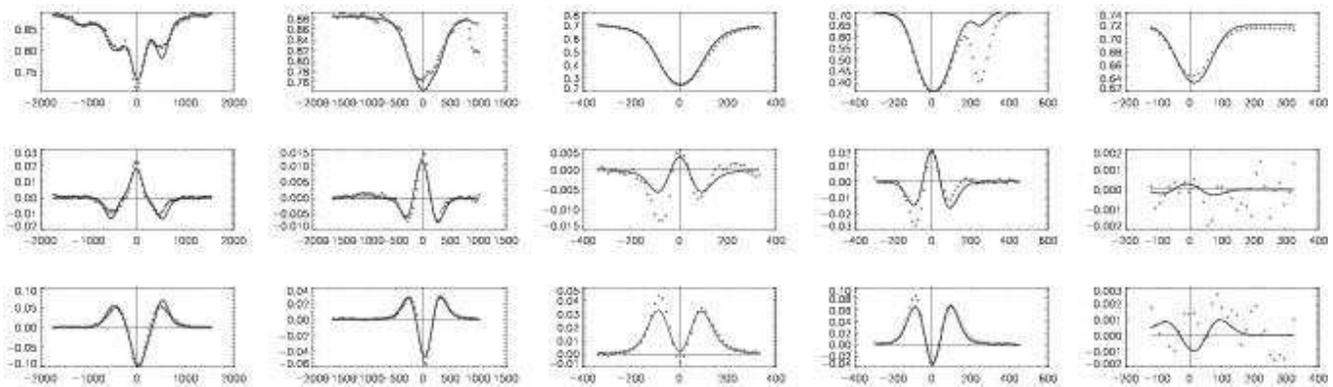}}}
\caption{Comparison of the observed (+) and best-fit profiles ({\em solid
    line}) for one spatial position in the neutral line of Stokes V. {\em Left to right}: infrared lines at 1564.8 nm
  and at 1565.2 nm; visible lines at 630.15 nm, 630.25 nm and 630.37 nm. {\em Top to bottom}: Stokes $IQUV$ as fraction of the continuum intensity. The black vertical lines mark the zero-wavelength of the respective
  wavelength scales. The dispersion on the x-axis is in m{\AA} for all
  spectra.  \label{invcomp}}
\end{figure*}
The data from TIP and POLIS were treated with the flatfielding
procedures and polarimetric corrections for instrumental effects \citep[see
for example][]{beck+etal2005b,beck+etal2005a}. Residual crosstalk between the different Stokes parameters was estimated to be on the order of $10^{-3} I_{\rm c}$. The remaining rms noise in the profiles  in continuum windows was $4 \times 10^{-4} I_{\rm c}$ for the IR spectra and $10^{-3} I_{\rm c}$ for the visible spectra. 

The data alignment was done in the same way as described in detail in \citet{beckthesis2006} or the Appendix of \citet{beck+etal2006d}. The wavelength scale was also set up like in the latter, with the blue-shift values predicted by the quiet Sun (QS) model of \citet{borrero+bellot2002} as a reference. The Stokes profiles of each spectral range were normalized to the continuum intensity of the quiet Sun at disk center in a two-step procedure. An average QS profile for each wavelength range was calculated from pixels located outside the sunspot. This profile was normalized to unity with its respective average continuum intensity. The off-center position was taken into account by a multiplication of the QS profile and all other profiles with the appropriate limb-darkening coefficient.
\section{\label{sec_inv}Inversion of the Stokes profiles}
The four visible and the three infrared lines have
been inverted together using the SIR code \citep[Stokes Inversion
based on Response functions;][]{cobo+toroiniesta1992,cobo1998}. To facilitate a good choice for the inversion setup and the initial
model components to be used, I created masks of the inversion type. I used
the intensity maps in infrared to define the boundaries between umbra and
penumbra, and penumbra and surrounding granulation, respectively. Outside the
sunspot, I used a threshold in polarization degree to distinguish between
field-free pixels and those with a polarization signal sufficiently large to
derive the magnetic field. The threshold was set to 0.4 \% for the infrared spectral lines (1564.8, 1565.2\thinspace nm) and 0.75 \% for the visible spectral lines (630.15, 630.25\thinspace nm). If any one of the lines exceeded its respective threshold, a magnetic field was assumed to be present.

All model atmospheres were prescribed as a function of continuum
optical depth, $\tau$, in the range from $\log \tau = 1$ to $-4$. For the
initial temperature stratification, I always adopted the HSRA
model \citep{gingerich+etal1971}. Temperature was allowed to be varied with two nodes,
i.e.~perturbations of the initial model atmosphere with a straight line of arbitrary
slope were possible. A contribution of stray light to the observed profiles was
also always allowed for; as proxy of the stray light profile the average QS
profile was used.

I employed a two-component inversion with one magnetic and one field-free component for pixels outside the sunspot with significant polarization signal. The free parameters of the
field-free component were the temperature, $T$, and line-of-sight (LOS)
velocity, $v$. For the magnetic atmosphere component, additional parameters were the field strength, $B$, the LOS inclination, $\gamma$, and the azimuth of the
field in the plane perpendicular to the LOS, $\psi$. Except for temperature, all  atmospheric parameters were assumed to be constant with optical
depth. Those spectra, whose polarization degree was below the threshold, were not inverted. I assumed a single magnetic component in the
umbra. For all pixels in the penumbra, the inversion setup used two independent magnetic components in each
pixel. The macroturbulent velocity was the only parameter that was forced
to be equal in both components. All quantities besides $T$ were again constant with depth. 
\begin{figure*}
\centerline{\resizebox{16.cm}{!}{\includegraphics{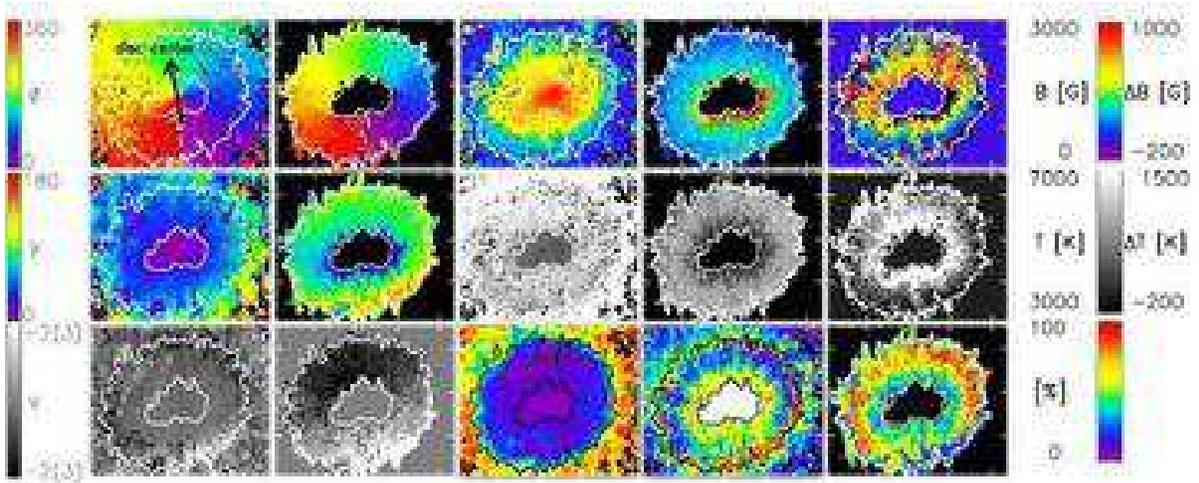}}}
\caption{{\em Top row, left to right}: magnetic field azimuth of the
  background component (bg), field azimuth of the flow channel component (fc),
  field strength of bg, field strength of fc, and difference of the field
  strengths  (bg - fc). {\em Middle row, left to right}: field inclination of
  bg and fc, temperature of bg and fc, difference of temperatures (bg -
    fc). {\em Bottom row, left to right}: LOS velocity of bg and fc, stray
  light contribution, filling fraction of bg and fc. Inclination and azimuth
  are given in the LRF frame; velocities are in kms$^{-1}$ with negative
  velocities pointing towards the observer. The velocity limits in parentheses
  refer to the flow channel component. White contour lines denote the inner
  and outer boundary of the penumbra. The black arrow in the azimuth map points towards disc center.\label{tempbfieldinv}\label{azirot}}
\end{figure*}

For SIR, the two inversion components are fully equivalent; the changes
applied to their initial model atmospheres are only driven by the need to
minimize the deviation between observed and synthetic profiles. In the
inversion of the spectra of neighboring pixels the roles of the two inversion
components thus may be exchanged, i.e.~component 1 may show a large flow
velocity and component 2 is at rest in one case, whereas on the next pixel it
is the opposite way. The results of the inversion have thus to be sorted
somehow to provide a smooth spatial variation. I used the inclination to the
surface as criterion to separate the two inversion components into different
maps: the more vertical component of each pixel is assumed to be the static
background (bg) field of the spot, and the more inclined component to be the
flow channels (fc). This selection by a single criterion can lead to
  ambiguities, when the inclination of the two components is similar. Near the
umbra-penumbra boundary, the field inclinations of the two components were
nearly identical for both observations of the sunspot
\citep[cf.][Fig.~5.14]{beckthesis2006}. In this area, I modified the criterion
and used the temperature of the components: the hotter component was chosen to
be the bg component, the cooler one the fc. This yielded smooth temperature
maps, which otherwise showed clear indications
that the identification of the components by inclination was wrong.

Figure \ref{invcomp} shows an example of observed and best-fit profiles for
one location in the neutral line of Stokes $V$. The weak \ion{Fe}{I} line at 630.35\thinspace nm has been left out in the graphs to save some space. The inversion is not able to reproduce asymmetric profiles, and thus fails to retrieve the observed
profiles down to the finest details (e.g., Stokes $Q$ and $U$ of the visible
lines). It however still catches the overall shape of the profiles quite well. Especially in Stokes $V$ one and the same atmosphere leads to strongly
differing profiles in, e.g., 1564.8\thinspace nm and 630.15\thinspace nm. The good agreement of observed and best-fit profiles in both wavelength
  ranges using constant field properties is related to the only small
  difference in formation height between the infrared and visible
  lines. \citet{cabrera+bellot+iniesta2005} studied the properties of several
  photospheric lines and concluded that the formation height
  difference between the 1.56 $\mu$ lines and those at 630 nm is smaller than
  100 km. The profiles shown in Fig.~\ref{invcomp} have a non-zero NCP; the
  NCP is not reproduced by the best-fit profiles. Even if the observed sunspot has a non-zero NCP with different behavior in infrared and visible spectral lines \citep{mueller+etal2006}, the main information
  contained in the spectra is at first the average properties of the
  magnetic field like field strength. The NCP then
  yields information on the variation along the LOS, but always around the average value. In \citet{beckthesis2006}, I found that the analysis of the same data
set using an uncombed inversion model yielded almost identical average field
properties, i.e.~field strength or field orientation were not
influenced by the choice of the inversion setup with or without reproduction
of the NCP. This does not come surprisingly, because the inversion with
constant magnetic field properties already reproduces the observed (complex)
spectra fairly well (cf.~Figs.~\ref{vneutral} and \ref{invcomp}).
\section{Results\label{sec_results}}
Prior to further analysis, the inversion results were transformed from the
LOS reference frame into the local reference frame (LRF). The LRF is defined
such that z corresponds to the surface normal and increases with height, while
the x-axis points from the center of the spot towards the disc center. The
180$^\circ$ azimuth ambiguity was resolved by assuming a radial field
orientation inside the sunspot, and thus always choosing the azimuth solution
closer to radial orientation. 

\subsection{2-D maps of field parameters}
Figure \ref{azirot} shows the resulting field azimuth and inclination to the
surface normal. Only the results of the observation on the 9th of August at
30$^\circ$ are shown here and in the following, as the other observation
yielded similar results. The two inversion components have been
  separated by the criteria described in the previous section. There is an
easy way to control if the selection is reasonable. The component selected
as ``flow'' channels should also exhibit significant flow velocities. And indeed this is the case:
the flow channel component shows LOS velocities up to some
kms$^{-1}$, whereas the background component is almost at rest
(cf.~bottom row of Fig.~\ref{azirot}). This gives confidence that the other quantities from the inversion can really be ascribed to ``flow channels'' and ``background component''. 

Figure \ref{tempbfieldinv} also shows that the background component is
stronger almost throughout the whole penumbra, whereas at the outer boundary
the field strength becomes comparable. The patch of increased field
  strength in the fc component (red color) below and right of the umbra near
  its boundary to the penumbra is co-spatial to the locations, where the field
  inclinations of the two components were nearly identical. The strong
  difference to the field strength further radially outwards could 
  indicate that in the spectra of these locations no clear signatures of two
  magnetic components were present. The temperature maps (at $\log \tau = 0$)
  in the middle row show the background component to be hotter than the flow
  channels all throughout the penumbra. The stray light contribution to the best-fit profiles increases from 10
\% in the umbra to 15 \% in the penumbra, and strongly increases at the outer
spot boundary. Inside the penumbra, flow channels and background component have a
comparable filling fraction of around 50 \%. 

Radially oriented structures \citep[``spines'',][]{lites+etal1993} can be seen on the center side in the background components' field strength, the flow channel temperature, and in the relative filling fraction of the background component. The spines start with locally enhanced field strength of the background component in the umbra, and maintain stronger and less inclined fields throughout the penumbra. The filling fraction of the flow channel component inside the spines is only 50 \%, compared to about 70 \% outside the spines. This indicates that the Evershed motion may be (at least partially) suppressed in the spine areas. Their continuum intensity is reduced mainly in the inner penumbra; in the outer part the spines are less prominent in intensity.
\subsection{Azimuthal averages}
\begin{figure}
\centerline{\resizebox{8.8cm}{!}{\includegraphics{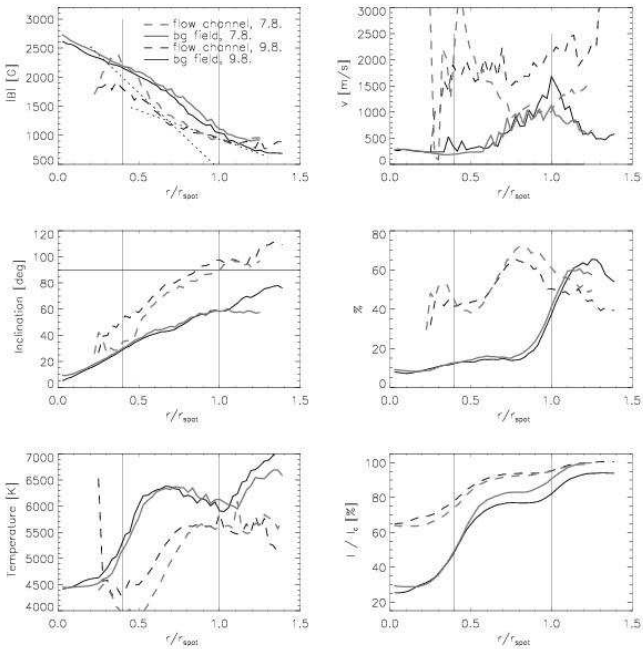}}}
\caption{{\em Left column, top to bottom:} Radial variation of B, LRF
  inclination,  temperature at log $\tau$ = 0. {\em Solid}: bg component, {\em
    dashed}: flow channel component. {\em Right column, top to bottom}:
  averaged absolute LOS velocity; stray light amount ({\em solid}) with
  overplotted filling fraction of the fc component ({\em dashed});
  intensity of visible ({\em solid}) and infrared continuum ({\em dashed}). In
  this case, black (grey) lines refer to the observation on 9th (7th) of August. The x-axis gives the fractional radial distance r/r$_{\rm spot}$, the black vertical lines mark the inner and outer boundary of the penumbra. The horizontal black line in the inclination plot marks 90$^\circ$. \label{radvarplot}}
\end{figure}

In order to investigate the mean properties of the two components, I averaged the field parameters azimuthally over ellipses with increasing radius. In the umbra the results of the one-component fit are used for the background component. In Fig.~\ref{radvarplot}, the results for the
azimuthal averages of both observations (7th and 9th) are shown for comparison. Starting in the lower right, the intensity in visible and infrared continuum
wavelengths has been used to place the transition from umbra to penumbra at
$r/r_{\rm spot}$=0.4, where $r_{\rm spot}$ is the spot radius of around 13
Mm. The contrast of umbra and penumbra is larger for the visible wavelength
range. The middle right panel shows that the relative filling fraction of bg
field and flow channels changes in the mid penumbra; the contribution of flow
channels increases from 50 \%  at $r/r_{\rm spot}$=0.65 to a maximum of around 70 \% at $r/r_{\rm spot}$=0.8. In the azimuthally averaged absolute LOS velocity the observation at
30$^\circ$ heliocentric angle shows a large difference of 2 kms$^{-1}$ between
bg component and flow channel, whereas at 6$^\circ$ the signature of the Evershed
flow is negligible. In both cases the bg component shows increasing
LOS velocities towards the outer boundary.

The field strength and field inclination of both observations are similar, indicating little change of the sunspot field topology in two
days. In both observations, the bg component is stronger by 0.5 kG in the
inner and middle penumbra, whereas at the outer boundary the strength of both components is nearly identical. The slope  with radius of the flow channel field strength
decreases in the mid penumbra at $r/r_{\rm spot}$ = 0.7 (dotted lines), but the field strength continues to drop with radius in both bg component and flow channels. The inclination of the flow channel component on average exceeds 90$^\circ$ for r/r$_{\rm spot} >$ 0.9, whereas the bg component never turns into horizontal fields. Its maximum inclination at the outer penumbral boundary is close to 60$^\circ$.

 The temperature plot in the lower left panel shows that the radial
  variation of the temperature in the background component follows closely the
  radial curve of intensity. The temperature curve of the flow channel
  component has a similar shape with reduced amplitude, but is displaced
  towards the outer penumbral boundary relative to the intensity or bg
  temperature curve. The decrease of the temperature in both components at the
  outer penumbral boundary is presumably connected to a trade-off between
  stray light and temperature in the inversion. At the outer boundary, the
intensity level and shape of the profiles allows to use a larger stray light
amount to reproduce the observed spectra, whereas in the umbra and penumbra
the QS profile simply does not fit to the spectra. These results are in good
agreement with the findings of \citet{borrero+etal2004} or
\citet{bellot+etal2004}, both in the absolute values of, e.g., field
  strength or field inclination, and in their radial variation. 

\begin{figure}
\centerline{\resizebox{5.5cm}{!}{\includegraphics{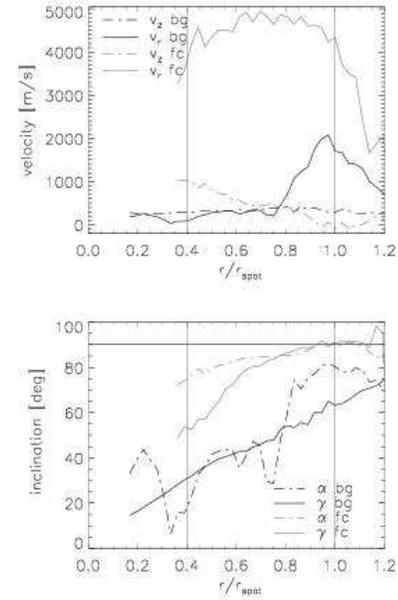}}}
\caption{{\em Top}: Horizontal and vertical velocities for both inversion
  components. {\em Black:} bg component. {\em Grey:} flow channel component. {\em
    Dashed:} vertical velocity. {\em Solid:} horizontal velocity. {\em
    Bottom}: field inclination, $\gamma$ ({\em solid}), and flow angle, $\alpha$ ({\em dashed}). \label{velcomp}}
\end{figure}
\subsection{Field-aligned flows}
With the assumption that on large spatial scales the flow velocities in the
penumbra are only due to the radially aligned Evershed flow, the LOS
velocity can be decomposed into its horizontal, $v_h(r)$, and vertical component, $v_z(r)$ \citep[e.g.][]{schliche+schmidt2000,bellot+balthasar+etal2003,bellot+etal2004}. This allows to derive the {\em flow angle}, $\alpha(r)$, i.e.~the inclination of the flow direction to the surface, at a given radius. The flow angle can be derived separately for each inversion component, and then be directly compared to the average field inclination of the component, $\gamma(r)$.  Figure \ref{velcomp} displays the corresponding results for the observation on 9th of August. The flow channel component shows upflows of around 1 kms$^{-1}$ ($v_z > 0$) in the
innermost penumbra, which turn into nearly horizontal flows with constant 4.6
kms$^{-1}$ all throughout the penumbra, and finally exhibit a small downflow
component for $r/r_{\rm spot} >$ 0.9. The background component shows a significant LOS velocity near the outer penumbral boundary, which could also be already
seen in the azimuthal averages of Fig.~\ref{radvarplot}. \citet{bellot+etal2004} have pointed out that this non-zero velocity of the background component is a necessary ingredient for correctly predicting the sign of the net circular polarization in the center side. The lower panel of Fig.~\ref{velcomp} displays that in the outer penumbra the flows in the flow channels are parallel to the field inclination to a high degree. In the inner penumbra and for the background component the agreement is worse, but this may also be due to the intrinsic inaccuracies of the method. The azimuthal fine structure of dark and bright filaments is ignored by assuming that the absolute flow velocity $v \equiv v(r)$ only depends on radius. Given these limitations, I think that the result supports the concept of field-aligned flows: upflows (downflows) are seen where the magnetic field inclination is below (above) 90$^\circ$.
\subsection{Signature of hot upflows in the mid penumbra\label{hotupfl}}
\begin{figure}
\centerline{\resizebox{6.cm}{!}{\includegraphics{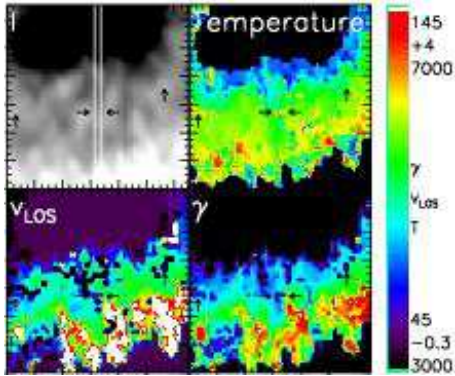}}}
\caption{({\em Top row}): continuum intensity and temperature of
  the flow channel component. ({\em Bottom row}): velocity and
  inclination of the flow channel component. The white lines mark the column, along which the parameters
  shown in Fig.~\ref{nextfig} were taken. Black arrows point towards hot upstreams in the mid penumbra.\label{prevfig}}
\end{figure}

\begin{figure*}
\sidecaption
\resizebox{11cm}{!}{\includegraphics{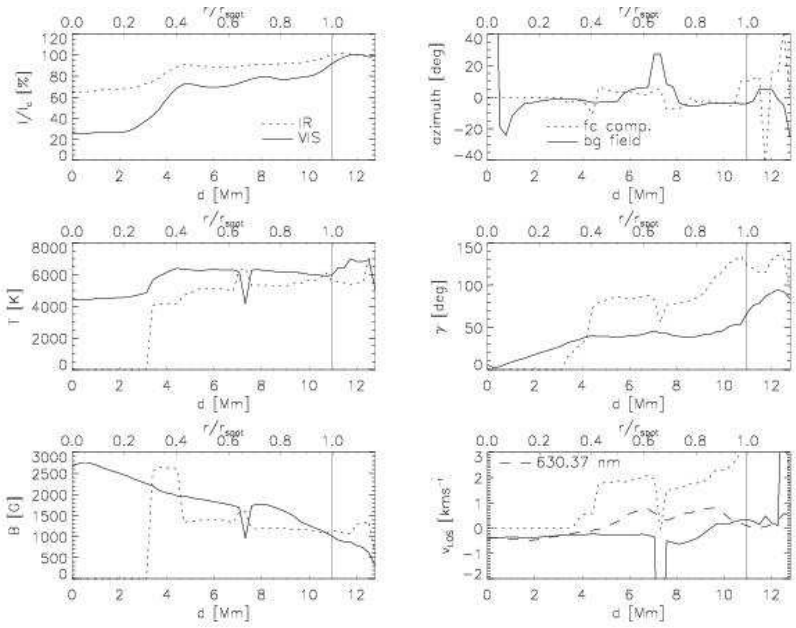}}
\caption{Parameters along the column marked in Fig.~\ref{prevfig}. {\em Left column, top to bottom}: continuum
  intensity, temperature, field strength.  {\em Right column, top to bottom}:
  field azimuth, field inclination, line-of-sight velocity. In all graphs
  besides intensity, {\em solid lines} indicate quantities of the background
  field and {\em dotted lines} those of the flow channel component. For
  intensity,  {\em solid} and {\em dotted} correspond to visible,
  respectively, infrared continuum intensity. The line-of-sight velocity of
  \ion{Ti}{I} is given by the {\em long-dashed} line in the velocity
  graph. The vertical black line marks the outer penumbral boundary. The
  second hot upstream intersected by the cut at $d$=6.2 Mm has a similar
  signature in all quantities as the upflows at the inner penumbral boundary.\label{nextfig}}
\end{figure*}

To investigate the radial structure of the fields without spatial averaging, I took the values of several quantities along a single column of the 2-D maps at around 0$^\circ$ field azimuth. Figure \ref{prevfig} shows this column and its
surroundings. At about the middle of the penumbra a hot upstream is
located. It has a strong signature in most quantities along the cut shown in
Fig.~\ref{nextfig}. The line-of-sight velocity in the flow channel component
drops from 2 kms$^{-1}$ further inwards to slightly below zero; it takes around 2 Mm to speed up again. The same is seen in the line core
velocity of \ion{Ti}{I}. The inclination of the field is reduced from
88$^\circ$ to 55$^\circ$, but jumps to around 77$^\circ$ on the next pixel ($\equiv$ 0\farcs37) already. \citet{rimmele2004} and \citet{rimmele+marino2006} also found a rapid change of upflows into horizontal outflows on scales of around 0\farcs6. The temperature of the flow channel component increases by around 1000 K on the upstream location. In the continuum intensity, an increase is seen near the
upstream, but displaced outwards by 1 Mm. Several similar hot upstreams in the
mid penumbra with a change of temperature, flow velocity and field inclination
can be seen in Fig.~\ref{prevfig}; two examples near the left and right border of the section shown are highlighted by arrows.
\paragraph{Significance of variation} The inversion procedure treats the
spectra of each position independently of their surroundings; the inversion
components are then separated with the two criteria (inclination,
  temperature) as described in Sect.~\ref{sec_inv}. This can produce some
  strong pixel-to-pixel variations in the 2-D maps of the inversion
  parameters. I think that for the case shown in Fig.~\ref{nextfig} I can
  exclude such an origin of the jump in atmospheric properties. The field
  inclinations differ significantly; the continuum intensities and the line
  core LOS velocity of the Ti line are quantities that are not related to the
  inversion procedure, and they both show also the signature of a hot
  upstream: a local intensity maximum with a vanishing LOS velocity. Note that
  the situation is different from the ``sea serpent'' shape suggested by
  \citet{schliche2002}, as I have no indications for downflows or field lines
  pointing down near the upstream. A possible configuration of the field lines
using the integration of the inclination (cf.~the next section) is shown in
Appendix \ref{appa}, Fig.~\ref{figg}.
\begin{figure*}
\sidecaption
\resizebox{10.cm}{!}{\includegraphics{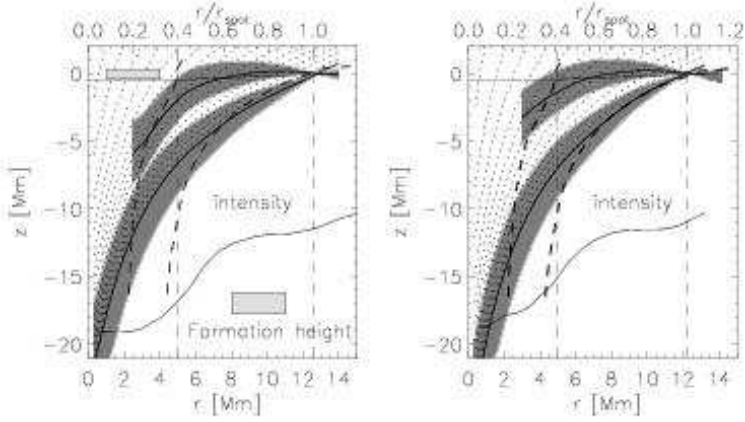}}
\caption{Integrated inclination of the data of 7th ({\em
    left}) and 9th ({\em right}) of August. ({\em Lower thick black line}): bg
  component; ({\em upper thick black line}): fc component. ({\em Dark grey
    shaded}): same curve including inclination variations. ({\em
    Thick dashed lines}): the boundary layers of the spot model of JS94. {\em Thin black line}: radial variation of the
  infrared continuum intensity. ({\em Vertical dashed lines}): inner and
  outer penumbral boundary. The {\em thin black line} at around z=0 km
  marks the location of the $\tau$ = 1 level. The {\em light grey shaded} area
  at (z=0 km,$r=3$ Mm) denotes the formation height(s) of the spectral
  lines. The {\em dotted lines} are identical to the bg component curve, but
  shifted in z to have the inclination observed at the $\tau=1$ level.\label{integ1}}
\end{figure*}
\section{Derivation of the 3-D field topology \label{sec_3dmodel}}
At first sight, the inversion with two magnetic components, whose properties are constant with optical depth, does not contain information on the vertical structure of the
magnetic fields. Taking into account that the information on the
field vector is available for an extended spatial region allows to derive a 3-D
model nonetheless by integrating the field inclination. The same method was already proposed in \citet{schliche+schmidt2000}, and applied to the radial variation of the flow angle derived from spectroscopy. It has also been used in \citet{solanki+etal2003a} and \citet{wiegelmann+etal2005} on measurements of the magnetic field vector. \citet{solanki+etal2003a} used the requirement that the field azimuth did not change along the integration direction, which is fulfilled for the sunspot observed.
\subsection{Integration of field inclination\label{integ}}
For each radial position, $r_i$, the azimuthally averaged inclination, $\gamma
(r_i)$, of background component and flux tube component are known
(cf.~Fig.~\ref{radvarplot}). With the assumption that the
inclination does not strongly change with depth below the surface (or height
above it), the geometrical height, $h$, of a field line in the solar
atmosphere can then be calculated from an integration of the inclination by
\begin{equation}
h(r_i) = \sum_{j=0...i} \frac{1}{\tan \gamma (r_j)} \cdot \Delta r \;, \label{heightdet}
\end{equation}  
where $\Delta r = 330$ km is the difference of the respective semi-major axes
of two subsequent ellipses.

Two additional steps were applied to the curves of $h(r_i)$ for background
field and flow channels. The flow channels are not observed in the
umbra; hence, no inclination values are available there. To have a common
reference height in both integrated curves, I forced the height of bg and fc at $r/r_{spot} = 1$ to be identical. This was motivated by the fact that
spectra at the outer white-light boundary still showed the signatures of multiple components. Thus, both flow channels and background component have to be present within the formation height of the spectral lines, and hence, must be present at the same geometrical height as well. I also assumed
that the isosurface of $\tau_{500} = 1$ has a slope of 3 deg when going from
the inner towards the outer penumbral boundary \citep{schliche+schmidt2000}, corresponding to a Wilson depression of 380 km in the umbra. The Wilson depression agrees with the number given by \citet{mathew+etal2004}, who however found no smooth  radial change, but a pronounced jump of 280 km at the umbral-penumbral boundary. Figure \ref{integ1} displays the finally resulting curves for both observations on 7th and 9th of August. Both observations yield very similar results.

The integrated inclination curve of the flow channel
component yields a slightly elevated arched loop. The apex height above the
$\tau = 1$ level is around 300-400 km. With the height fixed to 0 km at
the spot radius, the curve intersects the $\tau = 1$ level at around
$r/r_{spot} =0.6$. Outside the sunspot, the field lines point
downwards. As the background component inclination never reaches
horizontal fields, the integration yields a much steeper curve than for the
flow channels. If taken at face value, the curve reaches a depth of 2 (4) Mm
at $r/r_{spot} =0.8 $(0.6), implying a thick penumbra, not a shallow surface
layer. For the background component fortunately a direct comparison with a
theoretical model is possible. I overplotted the boundary layers between
umbra and penumbra and between penumbra and surroundings from the
magneto-static sunspot model of \citet[][JS94]{jahn+schmidt1994} in
Fig.~\ref{integ1}. I reduced the radius in their original calculation to 93
\% and 95 \%, respectively, of its value, to fit with the dimensions of the
sunspot in the observations, and shifted the curves in height to be at z=0 km at $r/r_{spot} = 1$ like the integrated inclination curves. For $r/r_{spot}$
between 0.6 and 1, the agreement between the integrated bg curve and the magneto-statical model is astonishingly good, whereas for smaller
radii the JS94 model is {\em steeper} than the integrated curve. 

The good agreement in the mid to outer penumbra comes a bit as a surprise: the
JS94 model gives the location (and thus also the field inclination) of the
boundary layer between spot and surroundings at depths up to some Mm, whereas
the integrated curve uses the inclination observed {\em at the surface} close
to the $\tau = 1$ level. The observed surface inclination thus seems to be
identical to the inclination in the deeper layers -- which was simply an
assumption in the derivation of the integrated curves. The deviation between
the curves also points in the correct direction: if the inclination changes
with depth, the fields should get more vertical in the deeper layers, as the
expansion of flux concentrations happens near the surface layer. Thus, the
surface inclination should be larger than in the deep layers, leading exactly
to the less steep integrated curve as obtained.
\paragraph{Limitations of the integration} Despite the surprisingly good
agreement with a theoretical model -- which also indicates that the main
assumption of a small depth dependence of the inclination could be valid -- the
integration of the surface inclination of course can not be fully correct. In
the left panel of Fig.~\ref{integ1}, I overplotted the approximate formation
height(s) of the observed spectral lines throughout the penumbra. All
information retrieved from the observed spectra thus only refers to this small
layer of the atmosphere. There is no guarantee that fields do not for
example bend strongly as soon as they leave the height range, in which the
spectral lines are sensitive. In fact, one expects exactly this behavior
for the field lines that pass the upper boundary of the formation height,
because of the exponential decrease of density with height. \citet{sanchezcuberes+etal2005} found no height dependence of the field inclination using spectral lines similar to the 630.15/25\thinspace nm pair, but this would be valid only inside approximately the same formation height range as plotted in Fig.~\ref{integ1}. To investigate the influence of inclination changes with height (or depth) on the integration of inclination, I repeated the integration with the
assumptions that the inclination may vary with height by $\pm$20 \% for the
background component, respectively, $\pm$15 \% for the flow channels. The upper
and lower limits of the retrieved curves are given by shaded areas in
Fig.~\ref{integ1}. As the expected change should be a deviation towards more
inclined fields in higher and less inclined fields in lower layers, any curve
that would not leave the shaded areas could be in agreement with the observed
surface inclinations, within the 15-20 \% range of variation allowed
for. Even if one takes the extreme case of a field line starting at
the lower and ending at the upper boundary of the shaded area, the
curves would not change enough to contradict the description above. The background component still would imply a thick penumbra; the flow
channels could change from non-elevated to strongly elevated loops, but
without giving rise to a new topology of the flow channels relative to the
background component.
\subsection{A three-dimensional model}
\begin{figure}
\centerline{\resizebox{8cm}{!}{\includegraphics{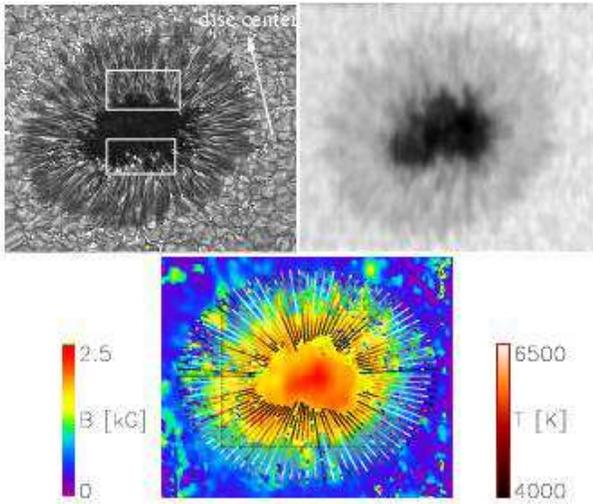}}}
\caption{{\em Top left}: Image of NOAA 10425 in the G-band from the DOT on La Palma,  taken on 9th of August about half an hour after the observations at the
  VTT. The white arrow points towards disc center. The white rectangles
    mark regions with an identically structured umbra-penumbra boundary in both
    intensity maps. {\em Top right}:  infrared continuum map. {\em Bottom}: Top view of the 3-D model. See text for the description.\label{penumbralgrains}}
\end{figure}

In the previous section, the integration was performed for azimuthal
averages, where the azimuthal fine-structure of the penumbra is lost. The integration can however also be performed for
individual radial cuts. To simplify the calculation and to smooth out slightly
the pixel-to-pixel variations in the inversion results, I decided to
use 92 bins\footnote{It needed to be a multiple of 4 for technical reasons.} of
about 4 deg angular extend for the construction of a 3-D model that takes the azimuthal fine-structure into account. The model itself is represented in the following way: the integration of the
background component gives a mesh of height with radial and azimuthal position,
$h(r_i, \phi_i)$. This mesh is interpolated to a smooth surface, whose color
is set to represent the field strength of the background component. The flow channels are overplotted as thin lines, with a color code
corresponding to their temperature. The temperature is scaled
individually between the respective minimum and maximum for each of the 92
bins; thus, the color bar only gives an average range of temperature. This description applies to Figs.~\ref{penumbralgrains} and \ref{integ2}.

As an intermediate step to the full 3-D model, Fig.~\ref{penumbralgrains}
shows a top view of the model. This view corresponds to the commonly
used 2-D maps of physical quantities. For comparison, I also show a
speckle-reconstructed image of NOAA 10425 in the G-band taken with the Dutch
Open Telescope (DOT) on 9th of August 2003 about half an hour later than the
observation analyzed here. The figure serves as a comparison of the
  spatial resolution of the polarimetric to speckle-reconstructed
  data. The boundary between umbra and penumbra has the same shape in both
  observations (cf.~inside the white rectangles), even with the time
  difference of half an hour. Bright penumbral grains can be seen in the DOT map near the umbral boundary, but only on the limb side.

Figure \ref{integ2} shows the full 3-D model for both observations
for three different viewing directions\footnote{Two animations of the model for the 9.8.2003 are available in the online material.}. I emphasize that the model is derived more or less
straightforward from the observed spectra, with in principle the single assumption that the integration of the field inclination is reasonable. If the model properties are projected to the surface at z=0 km, one has exactly the vector field and
thermodynamic properties that gave the best-fit to the observed Stokes
profiles in -- including the weak line blends -- 3 infrared and 4 visible spectral lines. The model(s) show some interesting peculiarities and give rise to several
interesting questions that could be addressed by analyzing them. The
observation close to disc center (7.8.2003, 6$^\circ$ heliocentric angle) is very uniform all around the spot, whereas in the other case
there is a much larger difference between center and limb side, where the flow
channels appear to be much more elevated. The background component exhibits a
ragged subsurface structure, where less inclined and steeper fields coincide
with an increased field strength. This could be closely related to the
question if these ``spines'' \citep{lites+etal1993} are co-spatial to darker or
brighter filaments. In general, a study on the relation of intensity to this
specific geometry would be interesting, because positive as well negative
correlations between field strength, inclination, and intensity have been
found up to now. Before an intense study on the question of the
validity of the integration of inclination has been performed, I refrain from
speculating further at this point. The 3-D models show some similarity to the artists sketch of a sunspot in Fig.~4 of \citet{weiss+etal2004}, but I remark that they are derived (directly) from observations here.
\begin{figure}
\centerline{\resizebox{8.cm}{!}{\includegraphics{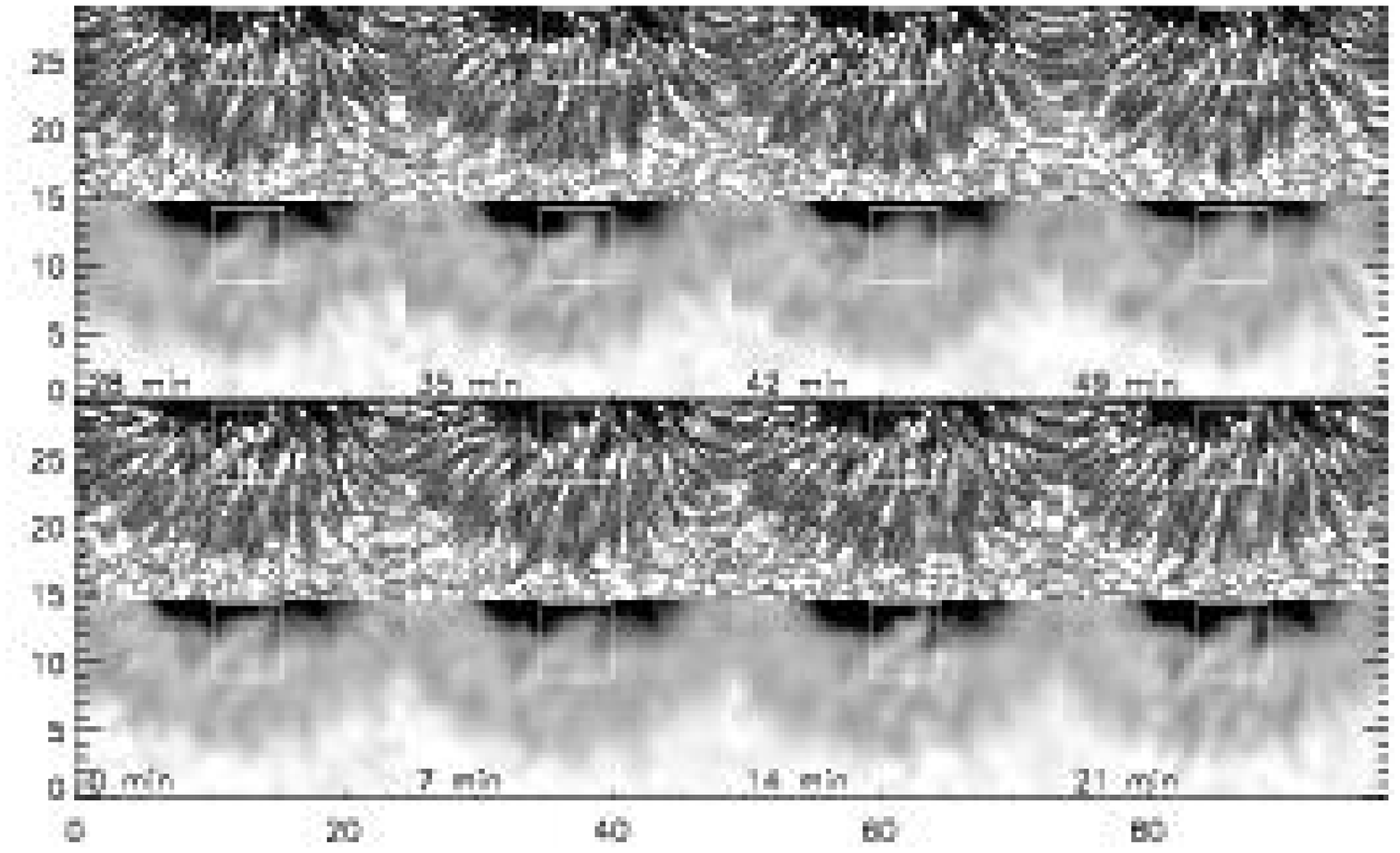}}}
\caption{The temporal evolution of the penumbra with 7 min
  cadence. Data of NOAA 10425 on 9th of August 2003, 9:36 to 10:40 UT. The
  lower row is the infrared continuum intensity, the upper row the co-temporal
  and co-spatial map in the G-band from the DOT. Tick marks are arcsec; time
  increases from left to right in each row.\label{teempevol}}
\end{figure}
\section{Temporal evolution \label{tempevol}}
To investigate the temporal evolution, I used a time-series taken after the
map on 9th of August. Figure \ref{teempevol} shows a small subsection of the
full field-of-view (FOV) at the DOT together with a co-temporal observation taken at
the VTT from 9:36 to 10:40 UT, after the map analyzed to derive the field topology. This is the upper part of the FOV containing the sunspot  which was not used in \citet{beck+etal2006d}, where the data properties and alignment method are discussed in more detail. The images show a peculiarity of the penumbral dynamics that I think has not been given enough
 attention up to now: the global structure of the sunspot, i.e., the boundary
 between umbra and penumbra marked with a white rectangle ,
 does not evolve at all in around 1 hour, whereas the intensity pattern of
 bright penumbral grains (PGs) evolves much faster on scales below 30 minutes
 (cf.~also Appendix \ref{appb}). This suggests that the dynamical
 evolution does not change the geometry. In the context of the previous
 sections, this could be interpreted as a static background component with a
 dynamic flow channel component. 
\section{Discussion\label{sec_disc}}
The penumbra has been subject of many studies, which in most cases did not
agree in their results. The observational findings are converging in
some points, which I think I can substantiate further by the results of this
investigation. The observations agree that the topology of the penumbra is
complex. The Evershed flow happens along the nearly horizontal flow channels,
whereas the average field inclination is {\em not} horizontal \citep[e.g.][]{bellot+etal2004}. The Stokes V
profiles of Figs.~\ref{vneutral} and \ref{invcomp} clearly show that at a
resolution of 1'' at least two independent magnetic components are present in
each pixel. This could be an artifact due to the spatial resolution, but if
the picture of the uncombed penumbra is valid, it will be the case {\em even
  if the flow channels were fully resolved}. As long as the flow channels are
not optically thick and located inside the formation height of a spectral
line, both horizontal and more vertical fields would be seen at the same
time. Stokes spectra with higher spatial resolution from, e.g., the recently
launched HINODE satellite could be used to see if the signature of multiple
components disappears at high spatial resolution. 

Some results turn out to be identical regardless of the inversion
method employed, being it a simple or complex model. If polarimetric profiles are analyzed in terms of two components (or also one component, but with gradients \citep{borrero+etal2004}), one retrieves a stronger, less inclined component
with small flow velocities, and a strongly inclined one, which harbors large
flows. The flow is located inside the magnetic fields, as the flow velocity is
derived from the Doppler shifts of the polarization signal. The same
classification into less inclined static field and more inclined flow channels
was obtained by \citet{langhans+etal2005} from magnetograms. All recent
inversions performed agree that the more inclined component bends downwards in
the outer penumbra
\citep{westendorp+etal2001,bellot+etal2004,borrero+etal2004}. Finally, hot
upflows were found at the inner penumbral boundary in also some other recent
studies
\citep{schliche+etal2004a,tritschler+etal2004,borrero+etal2005,rimmele+marino2006,bellot+etal2006}.

With respect to the topology of the fields, I introduced the concept of the
integration of the surface inclination. This approach relies on the assumption
that the field inclination does not, or only slightly, depend on depth or height
in the solar atmosphere. Even if the validity of the assumption has not been
addressed at all in this work, the method yields  reasonable
results. For the background component, I find a surprisingly good
agreement of the integrated curve with the outer boundary layer between the
sunspot and its surroundings in the magneto-static model of JS94. This is especially unexpected, as their boundary
line gives the field inclination in deep layers of some Mm, whereas I used
the inclination on the surface. The integration of the flow channel
inclination yields slightly elevated arched loops.
\begin{figure}
\centerline{\resizebox{4.5cm}{!}{\includegraphics{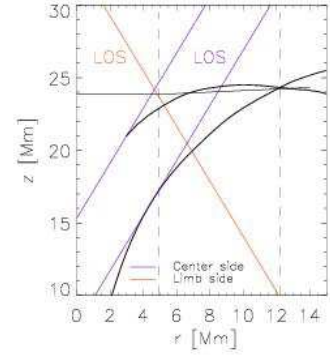}}}
\caption{The effects of the inclined line of sight on observations of the limb
  or center side. {\em Thick black}: integrated LRF inclination of the
  2-component inversion as in Fig.~\ref{integ1}, flow channel component (upper
  line) and background component (lower line). The LOS is overplotted in purple
  for the center side and in orange for the limb side.\label{effprojec}}
\end{figure}

Another peculiarity refers to the behavior of the inversion results at around
$r/r_{spot}$ = 0.65. At this location, the fill factor of the flow channels
strongly increases. \citet{westendorp+etal2001} found a local maximum of intensity at $r/r_{spot}$ = 0.6-0.7; \citet{bellot+etal2004} found a change of inversion results in the mid penumbra, a jump in the quantities of their more vertical component and an increasing filling fraction of the flow channel component. Interestingly, an intersection point of the surface with
the integrated curve is located near that radius. The inversion is performed
pixel-by-pixel on only the spectra from a single location, whereas the
integration uses the inclination results along a radial cut to create the curve
of the field lines. As these two procedures are fully independent of each
other, I think that this co-spatiality is not only pure coincidence. Taking
the integrated curve at face value, the obvious explanation is that flow
channels on average cross the surface around $r/r_{spot}$ = 0.65 and then have
a stronger signature in the observed spectra.
\subsection{Line-of-sight effects\label{losseff}}
The 3-D models derived by the integration for different heliocentric angles
show some deviations in their structure, even if azimuthally averaged values
are nearly identical. The model of the observation near disc center is very
uniform, while the other observation shows strong differences between limb and
center side. As both observations were analyzed in exactly the same way, the
difference should be contained already in the spectra. A second point related
to LOS effects is the comparison with a high-resolution image of the same
sunspot from the DOT telescope (cf.~Fig.~\ref{penumbralgrains}). Bright
penumbral grains can only be seen on the limb side.  This could be due to the simple geometrical effect depicted in Fig.~\ref{effprojec}. If the LOS is inclined to the surface, on the center side a limiting angle exists, where a structure will be hidden by its own continuation. The two LOS lines on the
  center side in Fig.~\ref{effprojec} intersect the bg and ft curves at locations where their inclination is the same as the heliocentric angle of the observation. The heliocentric angle of the observations on 9th of August was around 30$^\circ$; the {\em minimum} field inclination of the flow channel component on the limb side was around 25$^\circ$ (azimuthal average around 40$^\circ$, cf.~Fig.~\ref{radvarplot}). Thus, the absence of the penumbral grains  on the center side could be easily explained by the assumption that they are blocked by their own continuation as optically thick structures.
 \begin{figure*}
\centerline{\resizebox{17cm}{!}{\includegraphics{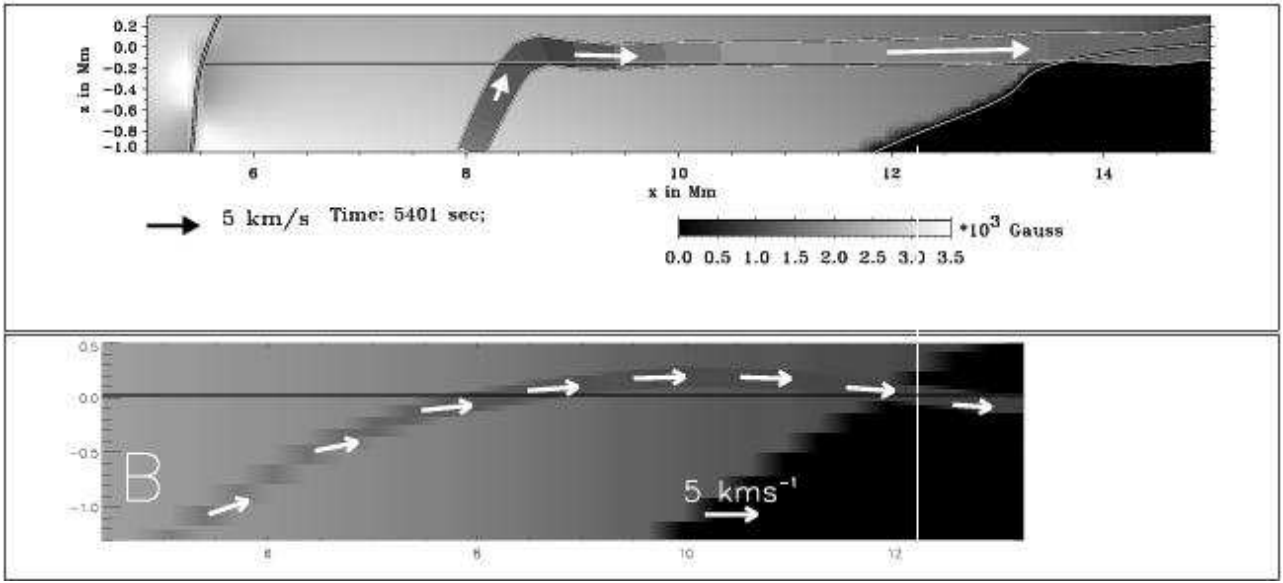}}}
\caption{Comparison of a snapshot of the Moving Tube Model ({\em top panel}, courtesy of R.Schlichenmaier)
  with the flow channels' geometry from the observation ({\em
    lower panel}). The plots have been arranged to have the same scaling and
  position in the horizontal axis which gives the distance from the spot
  center in Mm. Both panels display the flow channel component and the
  background component with their respective field strength as color
  coding. The color bar gives the color coding for field strength in both
    graphs. Velocities are overplotted as arrows, whose length gives the absolute velocity. The direction of the arrows for the lower plot is the field inclination.\label{geomcomp}}
\end{figure*}
\subsection{The Moving Tube Model}
From the observational point of view, a penumbral model of interlocked
horizontal and more vertical fields as given by the picture of the uncombed
penumbra is in good agreement with observed spectra, with their general shape
down to more subtle details like the net circular polarization of Stokes V
profiles \citep{mueller+etal2006}. From the various theoretical approaches,
the Moving Tube Model (MTM) of \citet{schliche+jahn+schmidt1998} is the only
one that results {\em by itself} in a similar geometry, including a flow along the horizontal channels. The MTM used the magneto-static sunspot model of JS94 as input for the
background component. I find that either the integrated inclination curve,
or, to be on the safe side, the observed inclination values of the background
field on the surface are in good agreement with the boundary layer between
sunspot and surroundings in the JS94 model. However, not only the
geometry of the background is in agreement with the observations. Figure
\ref{geomcomp} shows a direct comparison of the geometry, field strength, and
absolute velocity of a snapshot of the MTM during the final steady-state
of the simulations, and the same result as derived from the observations. 
For the location of the flow channel, I took the value of the integrated
inclination curve; for the outer sunspot boundary, I used the integrated
  background inclination curve. The flow velocity, and the field strength in
flow channel and background component were taken from the azimuthal averages
of the inversion results. The only quantity that was defined ad hoc was the
width of the flow channels, on which the present inversion does not yield any
information; I used a diameter of 200 km for it \citep[cf.~][]{beckthesis2006}. 
I find quite some similarity between the MTM and the results derived from the
observations.  Both field topology and the local properties (field
  strength, flow velocities in the inner and mid penumbra) do match reasonably well. Note that from the theoretical side the differences between the MTM and the integrated curve in, e.g., the height of the flow channel ($\sim$ 100-200 km) will lead people to argue that the curve suggested by the integration is inconsistent with any physics. I remind however that it reflects the azimuthal average of, at maximum, half-resolved flow channels, converted to a geometric height without much sophistication, whereas the MTM is supposed to describe an individual resolved structure. \citet{spruit+scharmer2006} and \citet{scharmer+spruit2006} raised the issue of the stability of flow channels of assumed circular diameter inside the penumbra, and claimed that no stable configurations are possible. The present investigation using model components without depth variation of the field properties actually does not yield information on this topic, but the inversion results and the integration of the inclination with the assumption of no depth variation would still apply for the case of flow sheets with a small lateral and large vertical extension. Finally, the agreement of the NCP predicted by the MTM and the observations is already partly discussed in \citet{mueller+etal2006}. A more detailed comparison of observed NCP, the NCP predicted from the MTM, and the NCP resulting from a fit of an uncombed penumbral model to the same observations used here is planned to appear in another paper.
\subsection{The ``gappy'' penumbra model}
\citet{spruit+scharmer2006} and \citet{scharmer+spruit2006} suggested a
  model of field-free gaps as a possible mechanism of the penumbral energy
  transport. Even if a direct comparison of their model with observations is
  difficult, as yet no (polarization) spectra were presented for their model,
  some of their arguments can be compared to the present findings.
  \citet{spruit+scharmer2006} suggested that the amount of stray light found
  in inversions could be
  related to the presence of a field-free component. The stray light amount in
the inversion of the present observations increases smoothly from around 10\% in the
umbra through the penumbra ($\sim$ 10-15 \%) to 80 \% at the outer penumbral
boundary. The 2-D map of the stray light contribution (cf.~Fig.~\ref{tempbfieldinv})
shows little spatial structure besides the radial increase. The stray light
inside telescope and the instruments TIP and POLIS was estimated to be around
15 \% \citep{reza+etal2007,cabrerasolana+etal2007}. There thus is little
indication for a field-free component inside of the sunspot in the inversion
results. 

 \citet{scharmer+spruit2006} argue that the inversion of polarimetric data has
 a high degree of ambiguity, as similar profiles can result from different
 atmosphere stratifications. Using the IR lines at 1.5 $\mu$m, I think that
 most of the ambiguities are removed. In \citet{beckthesis2006} and \citet{beck+etal2006d} 
 I investigated the influence of the spectral lines on the parameters
 retrieved by the inversion. I found that e.g.~field strength is restricted by
 the splitting of the 1564.8 nm line within a limit of around $\pm 100$ G. Most
 average quantities of the field topology (field strength, field orientation)
 are more or less uniquely restricted by the spectra; the main source of
 error are actually the spectra themselves: spatial resolution,
 signal-to-noise ratio, polarimetric sensitivity, and the polarimetric
 calibration. The question that however remains open is the 3-D organization
 of the magnetic fields. The present inversion yields field strength and
 orientation inside the formation height of the spectral lines, but is not able
 to differentiate between a vertical or horizontal interweavement of field
 lines. The approach of the integration of the field inclination assumes 
 coherent structures from one pixel to the next in the radial direction,
 which is in my opinion highly probable, but need not be the case.

The most prominent spectral feature inside the penumbra, the Evershed effect,
is not present in the gappy model. I remark that all velocities derived
here (besides the line core velocity of \ion{Ti}{I}) always refer to
velocities \emph{inside} magnetic fields. To create the multi-lobed profiles
in the neutral line of Stokes V, two components of magnetic fields with
different orientation and bulk velocities are needed. If the gappy model
solves the penumbral heat transport problems, still an explanation for the
Evershed flow is needed. The inclination of the bg field shows an azimuthal variation that leads to a
``gappy'' structure (cf.~Figs.~\ref{penumbralgrains} and
\ref{integ2}). However, the spatial scale of the variation is larger than
that predicted by \citet{scharmer+spruit2006}. If the integrated curves are
taken at face value, this also happens in layers far below the surface layer
of $\tau$ = 1. 

Another argument in favor of the gappy model is that the spatial
resolution of the observations I used may not be sufficient to detect the
signatures of the structuring suggested by Scharmer \& Spruit. In the other
direction, it should be possible to construct a sunspot model based on their
suggestions, calculate the resulting spectra in the 1.5 $\mu$m and 630 nm
lines, reduce the spatial resolution to around 1$^{\prime\prime}$, and then
invert the spectra with two depth-independent magnetic components. The results for the bg component could be compared with the present inversion results. Contrary to the regrettable sentence of \citet{scharmer+spruit2006} that ``\emph{the agreement with observations obtained with such [uncombed] models is of unquantifiable significance}'', I think it necessary to show that their model successfully reproduces spectroscopic or spectropolarimetric observations to support its validity.
\subsection{A model for the penumbral energy transport}
The long lifetimes (on the order of 1 hr) for filaments \citep[e.g.][]{langhans+etal2005}, and the lack of submerging flow channels led \citet{schliche+solanki2003} to the conclusion that interchange convection by rising hot flow channels is ``{\em not a viable heating mechanism}'' for the penumbra. This claim has been renewed recently by \citet{spruit+scharmer2006} and \citet{scharmer+spruit2006}, who criticize the ``{\em paradigm}'' of embedded flow channels and suggest that the penumbral fine-structure can be explained by a model of field-free gaps reaching almost up to the solar surface. The energy transport in their model is then effected by convection in the field-free plasma below the sunspot.

On the one hand, the findings of Sect.~\ref{tempevol} support the static
behavior of the sunspot fields: the shape of the umbral-penumbral boundary and
some especially dark patches inside the penumbra stay the same during around 1
hour. On the other hand, the intensity pattern of, e.g., the penumbral grains
is completely changed after less than half an hour (cf.~Appendix
  \ref{appb}). The time scales in the MTM model are of a comparable order. The snapshot shown in Fig.~\ref{geomcomp} was taken at 1 1/2 hour after the start
  of the simulation, but reflects the final steady-state solution. The flow
  channel in the MTM evolves rapidly in the beginning, and spans around half
  of the penumbra after 30 min \citep{schliche+jahn+schmidt1998}.

\citet{langhans+etal2005} used LOS magnetograms of the center side penumbra. These LOS magnetograms are not suitable to trace the dynamic evolution of the penumbra. On the center side, the field of the background component is parallel to the line-of-sight, whereas the dynamic flow channels are strongly inclined to it. Thus, the life times measured there reflect the slow evolution of the background component. The same applies to the findings of Sect.~\ref{tempevol}: the stronger field component will dominate the topology, and thus, the small amount of change in the geometry seen in Fig.~\ref{teempevol} implies that the {\em background component} does evolve only slowly. 
 
The conclusion on the impossibility of interchange convection due to the lack
of submerging flow channels needs some more explanations. To be convinced by
the argumentation, one would have to agree on the fact that the penumbra is
deep (some Mm), that flow channels originate from flux located initially on the
boundary layer between the sunspot and its surroundings, and that this flux
can become buoyant by heat input from the fully convective surroundings
outside the spot\footnote{I guess, less than half of the solar physicists will
  agree on that.}. The first point is suggested by the observations, while the
latter two are mainly based on the MTM. If one can agree on the
ingredients  above, I believe the penumbral heat transport
can be effected by hot rising flux tubes in full agreement with the
observations. If flux becomes buoyant at the outer sunspot boundary due to heating, this necessarily is a {\em repetitive} process. As soon as the hot flux bundle has risen from the boundary layer, new different flux will form the
boundary layer. This new flux would come from -- in the terminology
used throughout this paper -- the background component. After a time
span on the order of 30 min it would also have to become buoyant, and follow
the previously risen flux upwards. Due to the depth of the penumbra, several
flow channels could be stacked on top of each other at the same time. The
inversion results along a single column (cf.~Sect.~\ref{hotupfl} and Appendix
\ref{appa1}) suggest exactly this configuration: two flow channels  are seen along a radial cut at the same time at different locations in the penumbra.

The question of penumbral heating then changes to the question if the penumbral energy losses can be replenished by more than one hot flow channel with a characteristic repetition time around 30 min. It has been shown by \citet{schliche+bruls1999} and \citet{schliche+solanki2003} that the final state of the MTM is not able
to supply the penumbral heat requirements. The final state of the MTM is
however also {\em not} in agreement with the temporal evolution of penumbral
fine-structure. Penumbral grains do not persist for hours, but fade away after
reaching the umbral boundary, or turn into umbral dots without a visible
  connection to the penumbra \citep[e.g.][]{sobotka+etal1995}. The MTM never reaches this point, probably  due to its boundary condition, which fix the lower foot point of the flow channel to
the outer boundary layer between spot and surroundings. If the source of heat
input is the convective surroundings outside the spot, it is easily conceivable
that a flow channel would start to disappear, if it detaches completely from
the boundary layer.

The repeated ascent of flow channels of course would on the long term move all
flux away from the boundary layer, leading to the disappearance of the
penumbra after a short time. To replace the flux, {\em no submerging flow
channels} are needed. The replacement of the boundary layer can be effected
by a gradual re-arrangement of the flux of the background component. The open
question then is if a flow channel can become part of the background component,
after reaching the umbra. This should not be impossible: the inclination
difference between flow channels and background component in the innermost
penumbra is around 20$^\circ$ only. Furthermore, the inner, more vertical
foot point of the flow channel will contribute to the magnetic field
pressure term in the inner penumbra, pushing other field lines of
  the bg field radially outwards towards the magnetopause. 
\begin{figure}
\centerline{\resizebox{7.cm}{!}{\includegraphics{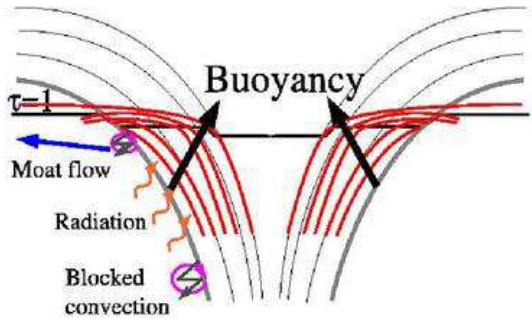}}}
\caption{Schematical model of the sunspot structure.\label{spotmodel}}
\end{figure}

\section{Summary \& Conclusions}
I have analyzed two spectropolarimetric observations of NOAA 10425, taken two days apart, at an heliocentric angle of 6$^\circ$ and 30$^\circ$, respectively. The observations consisted of Stokes vector polarimetry of four
visible spectral lines around 630\thinspace nm and three infrared spectral
lines at 1.5$\mu$m, taken with POLIS and TIP. I inverted the spectra with the
SIR code, using two independent magnetic model components. The field
properties were assumed to be constant with optical depth. I sorted the two
magnetic components by their inclination to the surface normal, and refer to
the more (less) inclined component as flow channels (background
component). In the innermost penumbra the inclinations of the two
  components were nearly identical, and I used the temperature as criterion instead.

The inversions of the same spot on the two days agree that the more inclined
component is weaker by around 0.5 kG in the innermost penumbra, but of same
strength as the more vertical background component at the outer spot boundary due to a less steep decrease of field strength with distance from the spot center,
like also found by \citet{borrero+etal2004} or \citet{bellot+etal2004}. I
find hot upstreams  in the mid penumbra, which appear in the more
inclined component.  The hot upflows show
  however more vertical fields in the flow channel component than on
  neighboring pixels. The more inclined field component shows high flow velocities up to
5 kms$^{-1}$ throughout the whole penumbra, which I find to be roughly aligned with the field direction. At the outer spot boundary, the flow channels on average bend slightly downwards to return to the surface (or submerge below it),
whereas the background component never exceeds an inclination of 60-70$^\circ$. Both inverted sunspot maps show a change of the relative filling fraction of
background component and flow channel component at around $r/r_{spot}$ = 0.65, where the filling fraction of the flow channel component increases.

To generate a geometrical model of the penumbral field topology, I integrated
the surface inclination in radial direction. This approach is valid, if the
field inclination (not field strength !) is independent of or only slightly
dependent on depth in the solar atmosphere. This method can on the one hand be used as
powerful visualization tool, as it allows to easily set geometry and all other
field properties in context. On the other hand, integrating separately the inclination
of background component and flow channel component, it allows to construct a 3-D
model of the sunspot that is basically the best-fit atmosphere model for the observed
profiles. In this 3-D model, the background component shows steeper ridges of
enhanced field strength, which maybe are identical to the ``spines'' of
\citet{lites+etal1993}. The azimuthally averaged flow channel component yields
arched loops that cross the $\tau$ = 1 surface at around $r/r_{spot}$ = 0.65,
where the filling fraction changes in favor of the flow channels. There is no
obvious reason, why these two things happen at the same location: the inversion is done pixel-by-pixel
without any information from neighboring pixels, whereas the integration uses
all inclination values in radial direction. I thus conclude that it is highly
probable that on average the flow channels {\it do cross} the surface at this
radius. 

In general, the inversion results are in agreement with the simulations of the
Moving Tube Model of \citet{schliche+jahn+schmidt1998} in several aspects (geometry of background component
and flow channels, radial variation of physical quantities). With a characteristic penumbral time scale of intensity variations below 30 min, and a depth of
the penumbra of some Mm as suggested by the integrated inclination, I think
that nothing found in the present observations would be in contradiction to a
penumbral heat transport by a series of hot, consecutively rising flow
channels, as depicted in Fig.~\ref{spotmodel}. An analysis of a time series of
sunspot observations using a two-component inversion and the integration of the field
inclination may be able to give a more direct proof if this scenario is
actually happening in the penumbra. I will attempt this task as next step using either the data shown in Fig.~\ref{teempevol} or observations from later times that were taken with the help of adaptive optics system (e.g., \citet{cabrerasolanaetal2006}).

\begin{acknowledgements}
The VTT is operated by the Kiepenheuer-Institut f\"ur Sonnenphysik (KIS) at the
Spanish Observatorio del Teide of the Instituto de Astrof\'{\i}sica de
Canarias (IAC). The DOT is operated by Utrecht University at the Spanish Observatorio del Roque de los Muchachos of the IAC. This work has been partly supported by the Deutsche Forschungsgemeinschaft under grant SCHL 512/2-1. The POLIS instrument is a joint development of the High Altitude Observatory (Boulder, USA) and the KIS. Discussions and advice from L.R.~Bellot Rubio, R.~Schlichenmaier, and especially the supervisor of my thesis at the KIS, W.Schmidt, are gratefully acknowledged. I thank the referee for pointing out the problems with the identification of the two components in the inner penumbra.
\end{acknowledgements}

\bibliographystyle{aa}
\bibliography{references_luis} 
\clearpage
\begin{appendix}
\section{Creation of multi-lobed profiles\label{prof_constr}}
\begin{figure}
\centerline{\resizebox{7cm}{!}{\includegraphics{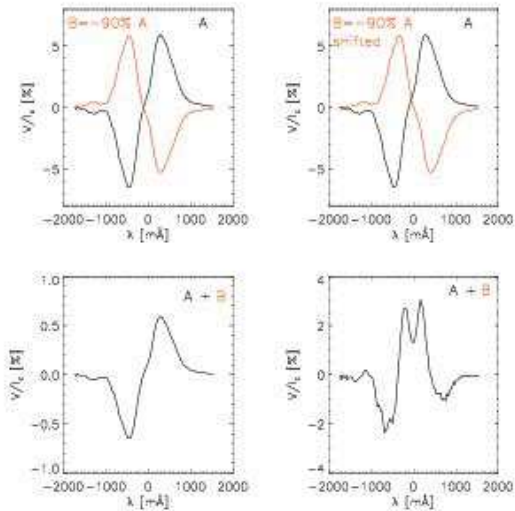}}}
\caption{A simple method to produce peculiar $V$ profiles. The addition of the two profiles A and B in the top row yields the profiles in the bottom row. {\em Left column}: profile B equals -90 \% of A and is located at the same wavelength. {\em Right column}: B equals -90 \% of A, but now has been shifted in wavelength. \label{neutralconstr}}
\end{figure}

The Stokes $V$ profiles in the neutral line on the limb side (Fig.~\ref{vneutral}) show a complex multi-lobed structure. Although the profiles strongly differ from those originating from a simple atmosphere with constant magnetic
field properties in the formation height of the spectral lines, only a few
ingredients are necessary to reproduce their shapes to first order. Figure \ref{neutralconstr} shows a simple experiment: the regular profile from the center side (rightmost graph of Fig.~\ref{vneutral}) is taken to be component
A; a second component B is constructed as  -90 \% of A. Addition of these two profiles generates again a regular two-lobed profile with reduced amplitude
(left column of Fig.~\ref{neutralconstr}). If component B is displaced in wavelength -- corresponding to the Doppler shift induced by a flow field of 2 kms$^{-1}$ -- the addition yields peculiar multi-lobed profiles like those observed in the neutral line of Stokes $V$. 
\section{2/3-D models\label{appa}}
\begin{figure}[ht!]
\centerline{\resizebox{8.cm}{!}{\includegraphics{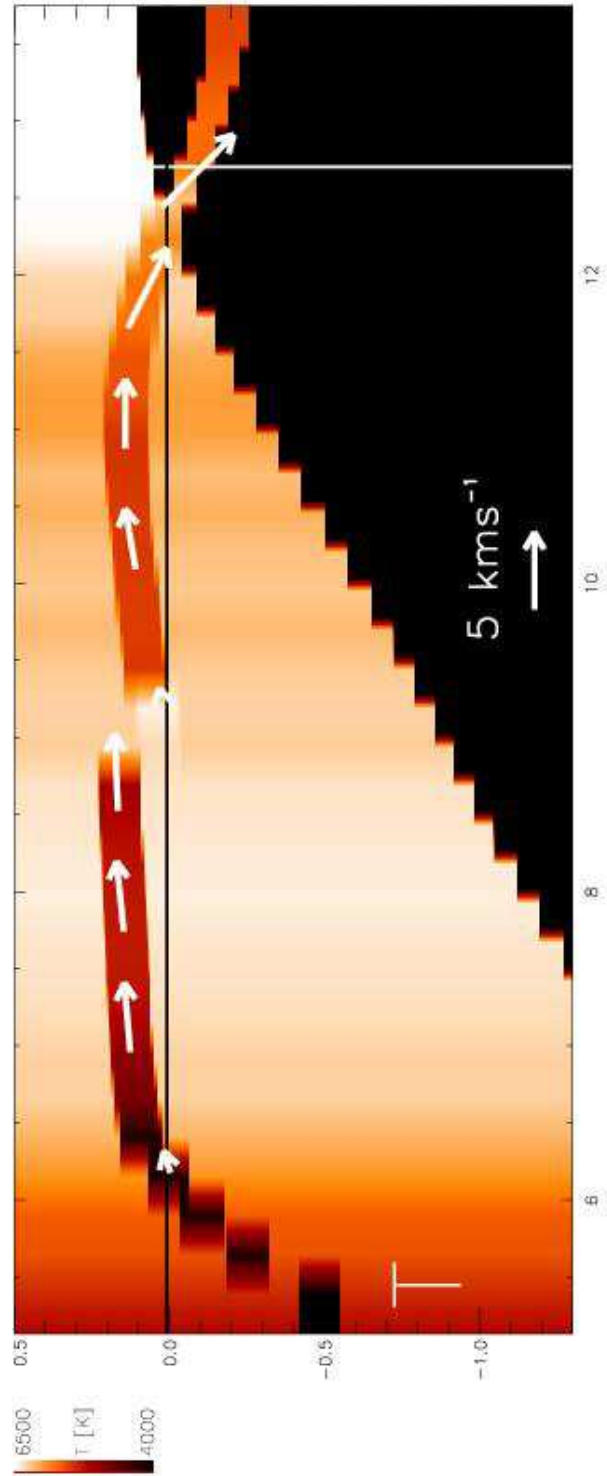}}}
\caption{The integrated inclination values along the single column marked in
  Fig.~\ref{prevfig}. Similar to Fig.~\ref{geomcomp}, but using temperature
  for the color coding. The integrated curve has been shifted down at x = 7
  Mm, where the hot upstream appears. \label{figg}}
\end{figure}
\subsection{2-D: Integration of single column\label{appa1}}
To follow the properties in a spatial cut without averaging, I took the
inversion results along a single column of the 2-D maps
(cf.~Sect.\ref{hotupfl}). The inclination of the cut can also be integrated in
radial direction. In Fig.~\ref{figg} I display the result in the same way
as in Fig.~\ref{geomcomp}, but this time with the temperature as color
coding. The  hot upstream is located at x = 9 Mm. The integrated
  curve was corrected to meet z=0 km at the outer sunspot boundary. Then I suggestively have shifted the integrated curve for x $>$ 9 Mm
  down by 150 km, to meet the $\tau$ = 1 level at  x = 9 Mm. As
the formation height covers some hundred kilometers, the geometry depicted
still would be in agreement with the observed spectra, even if of course I
have no proof that the second channel is below the first one.
\subsection{3-D models}
\begin{figure*}
\centerline{\resizebox{17.cm}{!}{\includegraphics{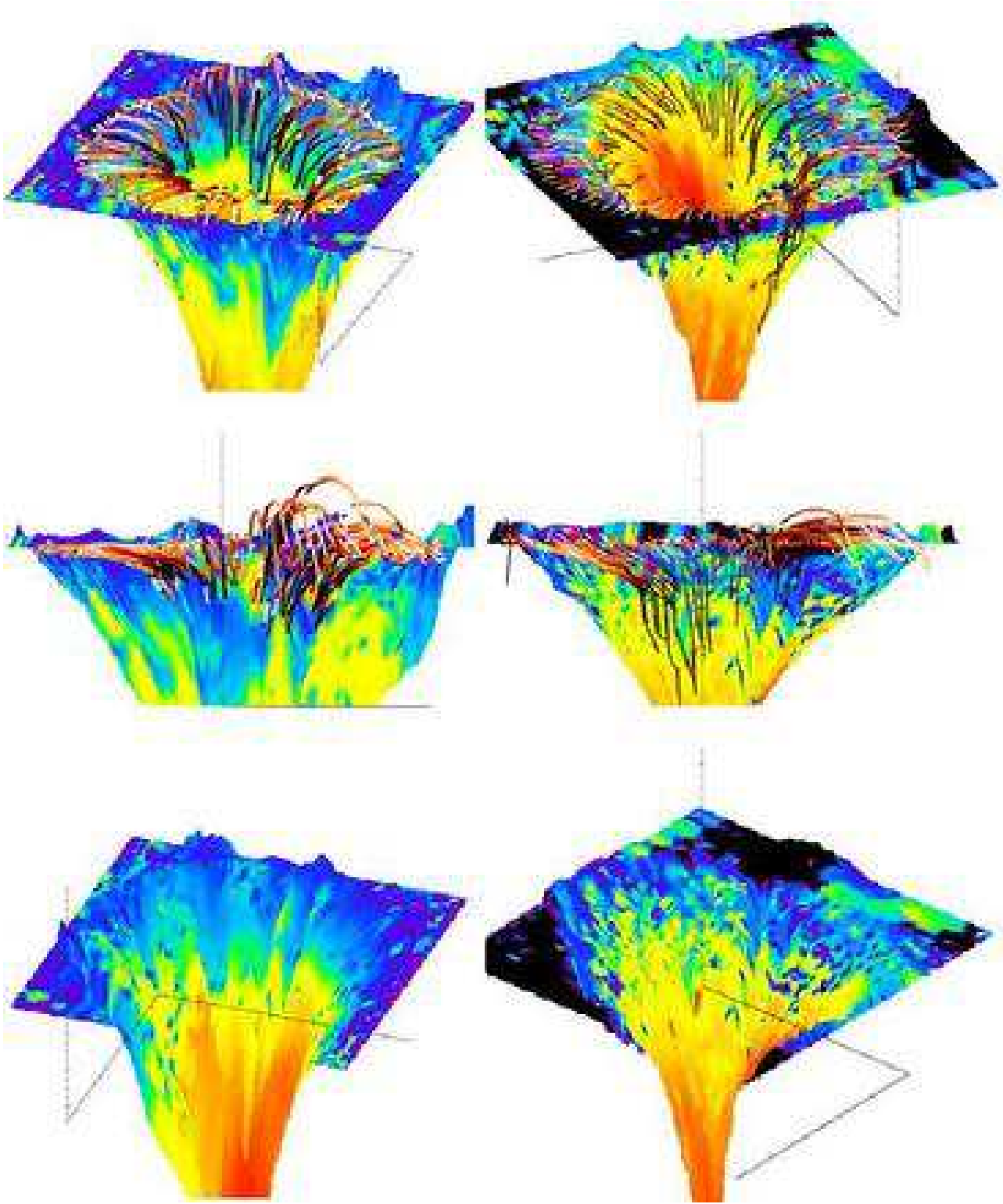}}}
\caption{The 3-D models from the integration of the inclinations to the surface
  averaged over 4 degree bins. ({\em Left column}): data of 9th of
  August. ({\em Right column}): data of 7th of August. ({\em Top to bottom}):
  view from above, side view, view from below. The color coding is identical
  to Fig.~\ref{penumbralgrains}. The plot of the flow channels was switched
  off for the view from below.\label{integ2}}
\end{figure*}

Figure \ref{integ2} shows the full 3-D model for both observations
from three different viewing directions: from above, the side, and slightly
below the surface. The color code and display method correspond to that of
Fig.~\ref{penumbralgrains}. Two animations showing the sunspot model of 9.8.2003 are available in the online material.

\section{Analysis of DOT time series\label{appb}}
\begin{figure}
\centerline{\resizebox{8cm}{!}{\includegraphics{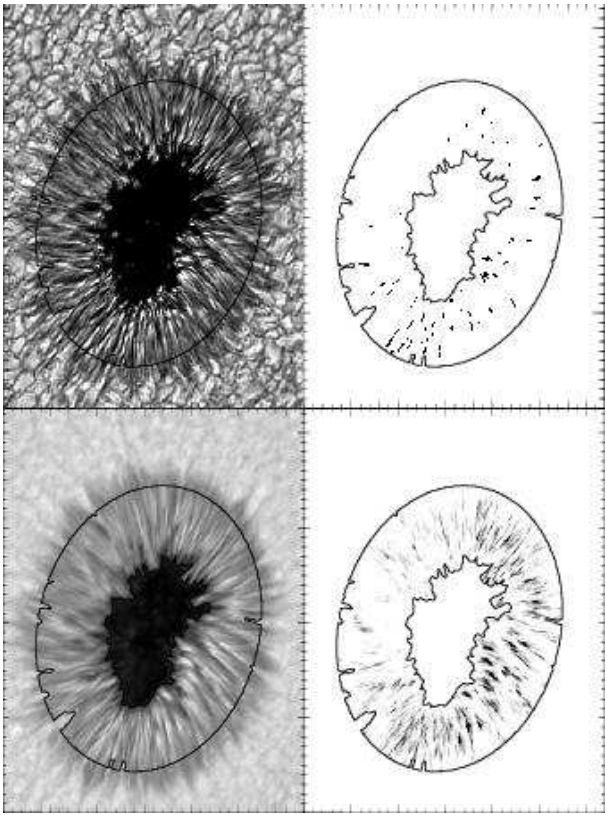}}}
\caption{Overview of the DOT data set. ({\em Top row}): single intensity image
  ({\em left}) with corresponding mask of brightenings ({\em right}). ({\em
    Bottom row}): Temporal averages over the time series. The black contours
  outline the area considered to be the penumbra. Tick marks are in arcsec.\label{tempaver}}
\end{figure}

To quantify the amount of dynamics in the penumbra, I used a 2 hours time
series of speckle-reconstructed images in the G-band from the DOT telescope,
taken on 9.8.2003 after the map used for the present investigation (30$^\circ$
heliocentric angle).
To derive the boundaries of the penumbra, I used the temporal average of the
time series (cf.~Fig.~\ref{tempaver}). To avoid counting brightenings at the
outer white-light boundary as penumbral grains (PGs), I further restricted the area to all points inside an
ellipse centered on the spot, but with smaller radius than the average
penumbra. I created a mask of all points inside the penumbra above a
threshold of 1.2 times the continuum intensity outside the spot for each of
the 214 images of the time series. The number and area of the PGs were taken
from the statistics of the individual masks.

\begin{figure}
\centerline{\resizebox{6cm}{!}{\includegraphics{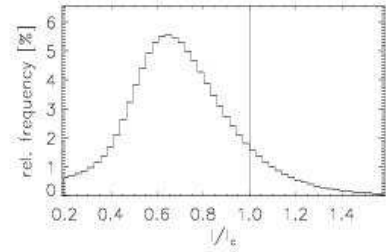}}}
\caption{Average histogram of penumbral intensities in the time series.\label{histint}}
\end{figure}

On a single radial cut starting at the center and ending at the outer
penumbral boundary, usually two or three individual PGs can be detected, even
if they not always all exceed the threshold used. The temporal average of the
masks clearly shows the dependence of the PGs on the azimuthal position inside
the spot: on the limb side, the PGs appear roundish and have a much higher
frequency than for azimuths of 90$^\circ$, respectively, 270$^\circ$, where
they usually appear as thin streaks. On the center side, they are almost
completely missing. As in a single image, usually two or more PGs can be seen
for any radial cut in the temporal average. The intensity of all points inside
the boundaries defined above during the time series yielded the histogram of
Fig.~\ref{histint}, which shows that around 10 \% of all points are on average brighter than the outside continuum intensity.
\begin{figure}
\centerline{\resizebox{8.cm}{!}{\includegraphics{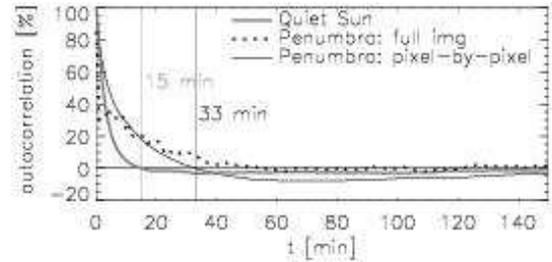}}}
\caption{Autocorrelation functions of the intensity in penumbra and a
  quiet Sun reference area.\label{autocor}}
\end{figure}

All following statistics and quantities have been derived from the full
time series in the area defined to belong to the penumbra. On average, in each image of the time series I found around 90
penumbral grains with an intensity above 1.2 times the continuum
intensity outside the spot. The average size of the PGs was 16 pixels, which
would correspond to a square area of 200x200 km$^2$. They covered an area
fraction of around 3 \% of the whole penumbra. Points with an intensity above
a continuum intensity of unity covered 10 \% of the full area. 

The intensity patterns in the penumbra disappear after a lifetime of around 30
min; the autocorrelation drops to zero around that time
(cf.~Fig.~\ref{autocor}). For comparison, I used a granulation area outside the
spot, which shows a characteristic correlation time of below 15 minutes. I
used two types of autocorrelation with the same result: autocorrelation of the
full images (restricted to the penumbra), and pixel-by-pixel autocorrelation,
where only the intensity of one spatial position with time was used in the
derivation of the autocorrelation.  The number of PGs is however constant to a
high degree. Thus, I conclude that the rate of the appearance of PGs should also have a characteristic timescale of around 30 min.

\end{appendix}
\end{document}